\documentclass[aps,prl,twocolumn,nopreprintnumbers,noeprint,superscriptaddress]{revtex4-1}
\usepackage{graphicx}
\usepackage{dcolumn}
\usepackage{bm}
\usepackage{amsmath}
\usepackage{amsfonts}
\usepackage{psfrag}
\usepackage{graphicx}
\usepackage{color} 
\usepackage{footmisc}
\usepackage{mathtools}
\usepackage{amssymb}
\usepackage[normalem]{ulem}

\newcommand\dout{\bgroup \markoverwith{\rule[0.2ex]{0.1pt}{0.4pt}\rule[0.8ex]{0.1pt}{0.4pt}}\ULon}

\begin{document}

\title{Excitons bound by photon exchange}

\affiliation{School of Physics and Astronomy, University of Southampton, Southampton, SO17 1BJ, United Kingdom}
\affiliation{Centre de Nanosciences et de Nanotechnologies (C2N), CNRS UMR 9001, Univ. Paris-Sud, Universit\'e Paris-Saclay, 91120 Palaiseau, France}
\affiliation{INO-CNR BEC Center and Dipartimento di Fisica, Universita di Trento, I-38123 Povo, Italy}
\affiliation{Laboratorio TASC, CNR-IOM, Area Science Park, S.S. 14km 163.5, Basovizza, I-34149 Trieste, Italy}
\affiliation{These authors contributed equally to this work}

\author{Erika Cortese}
\affiliation{School of Physics and Astronomy, University of Southampton, Southampton, SO17 1BJ, United Kingdom}
\affiliation{These authors contributed equally to this work}

\author{Linh Tran}
\affiliation{Centre de Nanosciences et de Nanotechnologies (C2N), CNRS UMR 9001, Univ. Paris-Sud, Universit\'e Paris-Saclay, 91120 Palaiseau, France}
\affiliation{These authors contributed equally to this work}

\author{Jean-Michel Manceau}
\affiliation{Centre de Nanosciences et de Nanotechnologies (C2N), CNRS UMR 9001, Univ. Paris-Sud, Universit\'e Paris-Saclay, 91120 Palaiseau, France}

\author{Adel Bousseksou}
\affiliation{Centre de Nanosciences et de Nanotechnologies (C2N), CNRS UMR 9001, Univ. Paris-Sud, Universit\'e Paris-Saclay, 91120 Palaiseau, France}

\author{Iacopo Carusotto}
\affiliation{INO-CNR BEC Center and Dipartimento di Fisica, Universita di Trento, I-38123 Povo, Italy}

\author{Giorgio Biasiol}
\affiliation{Laboratorio TASC, CNR-IOM, Area Science Park, S.S. 14km 163.5, Basovizza, I-34149 Trieste, Italy}

\author{Raffaele Colombelli}
\email{E-mail: raffaele.colombelli@u-psud.fr}
\affiliation{Centre de Nanosciences et de Nanotechnologies (C2N), CNRS UMR 9001, Univ. Paris-Sud, Universit\'e Paris-Saclay, 91120 Palaiseau, France}

\author{Simone \surname{De Liberato}}
\email{E-mail: S.De-Liberato@soton.ac.uk}
\affiliation{School of Physics and Astronomy, University of Southampton, Southampton, SO17 1BJ, United Kingdom}

\maketitle

{\bf 
In contrast to interband excitons in undoped quantum wells, doped quantum wells do not display
sharp resonances due to excitonic bound states. In these systems the effective Coulomb interaction between electrons and holes typically only leads to a depolarization shift of the single-electron intersubband transitions ~\cite{Nikonov1997}.
Non-perturbative light-matter interaction in solid-state devices has been investigated as a pathway to tune optoelectronic properties of materials ~\cite{Kockum2019,Forn-Diaz2019}.  
A recent theoretical work ~\cite{Cortese2019} predicted that, when the doped quantum wells are embedded in a photonic cavity, emission-reabsorption processes of
cavity photons can generate an \textbf{\emph{effective}} attractive interaction which binds electrons and holes together, leading to the creation of an intraband bound exciton. 
Spectroscopically, this bound state manifests itself as a novel discrete resonance which appears below the ionisation threshold only when the coupling between light and matter is increased above a critical value. 
Here we report the first experimental observation of such a bound state using doped GaAs/AlGaAs quantum wells embedded in metal-metal resonators whose confinement is high enough to permit operation in strong coupling.
Our result provides the first evidence of bound states of charged particles kept together not by Coulomb interaction, but by the exchange of transverse photons.
Light-matter coupling can thus be used as a novel tool in quantum material engineering, tuning electronic properties of semiconductor heterostructures beyond those permitted by mere crystal structures, with direct applications to mid-infrared optoelectronics.}

\begin{figure}[htbp!]
\begin{center}
\includegraphics[width=8cm]{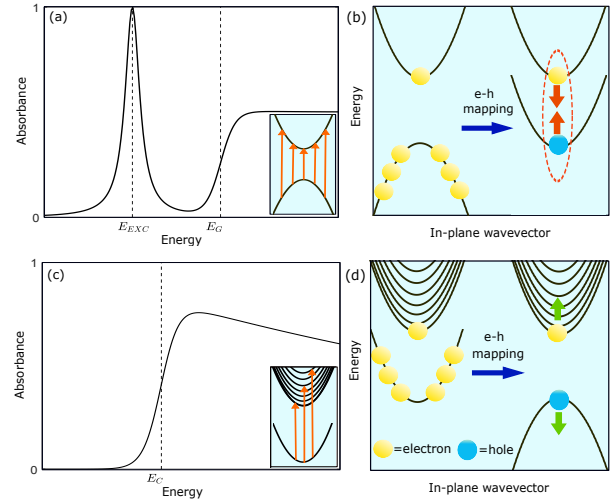}
\caption{\label{Figure_ISBT}
{\bf Coulomb effect in doped and undoped quantum wells.} 
In panel (a) we sketch the interband absorbance of an undoped semiconductor QW, dominated by an excitonic peak below the bandgap and a continuum absorption above it. In the inset we schematically illustrate the origin of the continuum part of the spectrum, understandable in a single-electron picture as interband transitions of electrons with different in-plane wavevectors.
In panel (b) we illustrate the standard electron-hole mapping, allowing us to describe a single electron vacancy in the valence band as a hole with positive charge and mass. We can thus understand the excitonic peak as an hydrogenoid bound state of an electron and a positively charged hole. 
In panel (c) we sketch instead the intersubband absorbance of a doped QW containing only one localized state, and a {\it continuum} of states above the barrier. Only the asymmetric, large bound-to-continuum absorbance is present, and no excitonic peak is visible~\cite{capasso1992}.
As shown in the inset the continuum is, in this case, due to each electron having multiple possible delocalised final states. 
The reason for the lack of an excitonic resonance is illustrated in panel d). As the initially filled electron subband has in this case a positive effective mass, the electron-hole mapping leads to a positively charged hole with negative effective mass, unable to form a bound state with the electron.
}
\end{center}
\end{figure}

Improvements in resonator design and fabrication have allowed to investigate the strong light-matter coupling regime in a multitude of solid-state devices ~\cite{Ballarini2019}. In such a regime the excitations of the system are not photons or matter resonances, but polaritons, hybrid quasiparticles which are linear superpositions of them.

If the light-matter coupling is large enough to couple multiple matter resonances, characterised by different spatial wavefunctions to the same photonic mode, the  ensuing polariton wavefunction will be a linear superposition of the uncoupled ones, possessing its own unique features, {\it a priori} different from those of the uncoupled modes. Such a phenomenon, referred to as {\it very} strong coupling ~\cite{Khurgin2001,Citrin2003,Khurgin2018} has only recently been experimentally demonstrated in inorganic microcavities, where it can be observed as a change of the exciton radius  ~\cite{Yang2015,Brodbeck2017}.

The question on whether such a mechanism can be pushed to the extreme, non-perturbatively modifying the excitation wavefunction and creating localised bound states from delocalised ones, has recently been theoretically investigated, leading to the prediction of discrete resonances appearing below the continuum ionization threshold for large enough values of the light-matter coupling strengths ~\cite{Cortese2019}.  The lower edge of the continuum corresponds to free electrons with no kinetic energy. The discrete resonances below such energy generated by the coupling with the photonic resonator have thus to be bound, as electrons have not enough energy to escape to infinity. 

The system investigated in Ref. ~\cite{Cortese2019}, microcavity embedded {\it n}-doped semiconductor quantum wells (QW), is particularly suitable to highlight this physics. First, it allows to achieve large light-matter coupling strengths ~\cite{Anappara2009,Todorov2010}, which can be tuned by changing the electronic density in the QW~\cite{Gunter2009}.
Second, it allows to design the bare electronic wavefunctions in order to have a single electronic subband trapped in the QWs, such that the bare electronic response does not present discrete intersubband transitions but only a continuum spectrum.
Third, in such a system the Coulomb interaction does not create bound states, which could otherwise hide the effect of the coupling to the microcavity. The lack of bound intraband excitons in doped QWs is well known ~\cite{Nikonov1997}. In Fig.~\ref{Figure_ISBT} we present a physical interpretation of such a phenomenon, useful to understand where the parallel with undoped QWs, whose optical response is instead dominated by bound excitonic states, breaks down.

In the case of an interband transition in an undoped QW the valence band, initially filled with electrons, has a negative effective mass. 
In the case instead of an {\it intersubband} transition in a doped QWs the same role is played by partially-filled conduction subband, which has instead a positive effective mass. Under the usual electron-hole mapping \cite{HakenBook} this leads to a positively charged hole with a negative effective mass. Such a quasiparticle moves in an external electric field as a negatively charged particle with a positive mass and it will thus be repelled by the electron. 
\begin{figure}[t]
\begin{center}
\includegraphics[width=8.5cm]{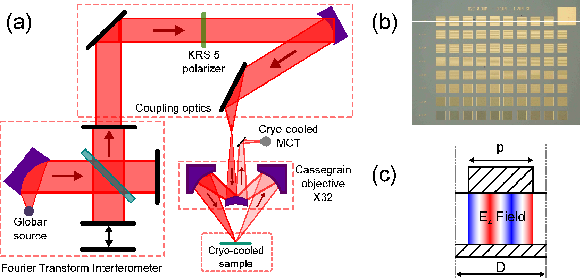}
\caption{\label{Figure_Experiment}
(a) Scheme of the experimental setup employed for the reflectivity measurements. It consists of a mid-IR microscope connected to an FTIR spectrometer. The source is a thermal globar lamp, and the detector is a liquid nitrogen cooled MCT. The incident beam is polarized normally to the metallic fingers, and the system operates in reflection. A compact, thin cryostat permits to perform the measurements down to a temperature of 78K.
(b) Optical microscope image of a typical sample set. Each device features a different finger width $p$. The top right device is a fully metallized surface that serves as mirror reference.
{(c) Distribution of the electric field component orthogonal to the metallic layers for one period of the structure and for the mode TM$_{02}$ of the ribbon resonator.
}}
\end{center}
\end{figure}

In order to provide a first evidence of photon-mediated bound states we realised a system similar to the one described in Ref. ~\cite{Cortese2019}, composed of $13$ GaAs/AlGaAs quantum wells embedded in highly-confining, grating-shaped gold microcavity resonators. They are 1D ribbons (or 1D patch cavities~~\cite{OPEXTodorov2010}), and the electromagnetic field, as sketched in Fig.~\ref{Figure_Experiment}(c), is almost completely confined below the metallic fingers.
The QWs are chosen thin enough to have a single trapped conduction subband, as the presence of a second one would lead to the creation of intersubband polaritons ~\cite{Dini2003}. Not only this would complicate the analysis of the data, but the presence of a bound-to-bound transition would almost saturate the available oscillator strength, leaving little for the bound-to-continuum transition we aim to measure.
For the purpose of checking this important point, we fabricated two samples, HM4229 and HM4230, differing in QW width and doping. Sample HM4229 contains 4-nm-thick GaAs QWs ($L_{\mathrm{QW}}=4$ nm), each doped at a density  $5\times 10^{12}$cm$^{-2}$, while sample HM4230 contains 3.5-nm-thick QWs ($L_{\mathrm{QW}}=3.5$ nm) doped at $4.77\times 10^{12}$cm$^{-2}$.

\begin{figure}[h]
\begin{center}
\includegraphics[width=8cm]{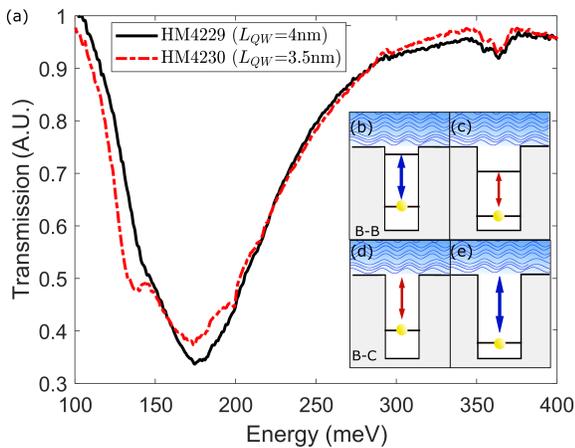}
\caption{\label{Figure_QW} {{\bf Bound-to-continuum nature of the transition.}} (a) The black solid line is the transmission measured at 300K from the sample HM4299, with $L_{\mathrm{QW}}=4$\ nm. The red dashed line represent instead sample HM4230, with $L_{\mathrm{QW}}=3.5$\ nm. In the inset we sketch how the transition frequency shifts as a function of the quantum well width for bound-to-bound (b,c) and bound-to-continuum (d,e) transitions.}
\end{center}
\end{figure}

Bound-to-bound and bound-to-continuum transitions undergo opposite frequency shifts when the quantum well width is reduced: the former ones blue-shift, while the latter ones red-shift. 
In Fig.~\ref{Figure_QW} we plot the transmission of the two samples before the gold deposition. 
\begin{figure*}[t!]
\begin{center}
\includegraphics[width=12cm]{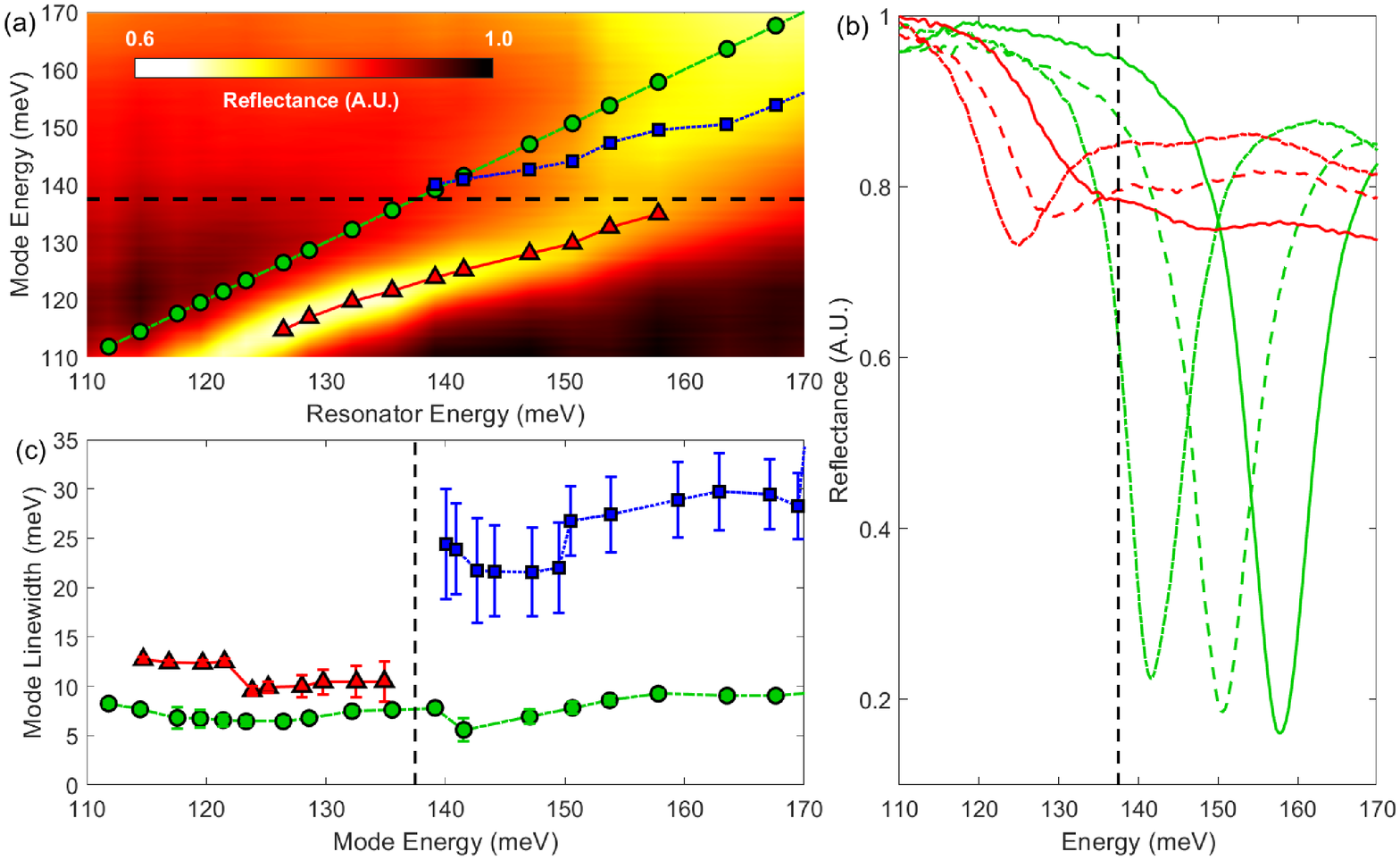}
\caption{\label{Figure_exciton} 
{\bf Experimental reflectivity data}
(a) Reflectance data from the doped sample HM4229 as a function of the resonator frequency. Superimposed we show the peak position extracted from the data, using
blue squares and red triangles for the datapoints respectively below and above the edge of the continuum. With green dot we mark instead data from the undoped sample taken as reference for the cavity energy. 
(b) Reflectance data for doped sample HM4229 (red) and the empty cavity (green) for resonator frequencies $\hbar\omega_c=157.8 \,\text{meV}$ (solid),  $\hbar\omega_c=147 \,\text{meV}$ (dashed),  and $\hbar\omega_c=141.5 \,\text{meV}$ (dot dashed).
(c) Linewidths for the different modes, as a function of the mode energy. 
In all the panels the dashed black line marks the edge of the continuum.}
\end{center}
\end{figure*}
We observe a very broad absorption that - being transverse magnetic polarized - stems from the doped QWs. We also observe some narrower feature around $140$ meV which can be identified as the edge of the continuum. Crucially, this feature does not blue-shift with decreasing QW width. It shows a transfer of spectral weight toward the red, while a bound-to-bound transition would have led to a blue-shift of the order of tens of meVs, proving the bound-to-continuum nature of the transitions in the bare QWs. 

Each sample was fabricated into metal-semiconductor-metal grating resonators with period $D$ and width of the metallic finger $p$, as sketched in Figs.~\ref{Figure_Experiment}(b) and (c), and following a procedure similar to the one in Ref.~\cite{Manceau2018}. 
The use of metal-metal resonators is important as it permits to couple to intersubband transitions, that are TM polarized, even at normal incidence.
In this case, however, the extreme thinness of the active region ($\approx\ 260\ nm$) forbids operation in the photonic-crystal regime~\cite{Chassagneux2008}, and places the system in the independent resonator regime instead. The resonance frequency $f$ is  set by $p$, and not by the period $D$, according to the simple formula
\begin{equation}
\frac{c}{f}= \frac{2n_{eff}p}{m},\ \ \  m\in\Bbb N,  
\end{equation}
with $c$ the speed of light.
As the electromagnetic field is extremely localized below the metallic fingers, the system essentially behaves as a Fabry-Perot cavity of length $p$~\cite{OPEXTodorov2010}. $n_{eff}$ is an effective index that takes into account the reflectivity phase at the metallic boundaries. 

We fabricated several grating-based devices, 200 x 200 $\mu$m$^{2}$ large, with $p$ ranging from 800 nm to 5 $\mu$m, in order to span a vast range of frequencies with the resonant cavity modes (see Fig.~\ref{Figure_Experiment}(b)).
We decided to employ the cavity mode with $m=3$, instead of the fundamental mode with $m=1$ that is typically employed in these systems, to simplify the fabrication procedure and increase the electromagnetic overlap factor. { This is evident from Fig.~\ref{Figure_Experiment}(c) that highlights three nodal lines for the resonant cavity mode.}

Each device is measured in reflectivity, at a temperature of 78K, using a FTIR microscope equipped with a very compact cryostat. A scheme of the experimental setup is shown in Fig.~\ref{Figure_Experiment}(a).
The main result of this work is reported in Fig.~\ref{Figure_exciton}, where in panel (a) we plot  the 78K reflectivity map of the HM4229 sample as a function of the bare resonator frequency. While above the ionization threshold, that is materialized by a black horizontal dashed line, we can observe an absorption continuum,
below it a narrow polaritonic resonance appears. It is red-shifted by more than 20 meV with respect to the bare resonator.
Overlaid on the colormap we plot the peak frequencies obtained by a multiple Lorentzian fit of the data.
Red triangles and blue squares are used respectively for frequencies below and above the identified ionization threshold marked by black dashed lines. 
For comparison, the green circles mark the frequency of the bare resonator, measured on an undoped sample (data in Supplemental Material). 
In panel c) the respective linewidths are shown as a function of the resonant mode frequency.

Below the ionization threshold it appears that the lifetime of the discrete polaritonic mode is mainly limited by the resonator lifetime. 
Above it, instead, we have a bound-to-continuum spectrum in which only very broadened and undefined features can be identified.
To better illustrate this finding, in Fig.~\ref{Figure_exciton}(b) we plot the reflectivity spectra of the doped sample (red curves) for different values of the bare resonator frequency, together with the resonator resonance measured on a undoped sample (green curves). From such a figure we can clearly observe how the doped sample develops a discrete resonance below the continuum edge, while an identical, but electromagnetically uncoupled sample presents none.
 
Note that in Fig.~\ref{Figure_exciton}(a) both the narrow discrete mode below the continuum edge (red) and the extracted peak of the broad continuum above it (blue) are at frequencies lower than the bare resonator (green). Moreover, they have very different linewidths which remain well separated for all cavity energies. These features are not compatible with a standard (bound-to-bound) polaritonic description, in which due to mode repulsion both the light and matter resonances are between the polaritonic modes. Moreover in a simple Hopfield model ~\cite{Hopfield1958} the coupled modes linewidth are weighted averages of the bare light and matter ones and they should thus cross across the anticrossing.
Data on sample HM4230 in strong coupling are reported in Supplemental Material. They are similar to the data presented here on HM4229 and point at the same conclusions.

\begin{figure}[htbp]
\begin{center}
\includegraphics[width=8cm]{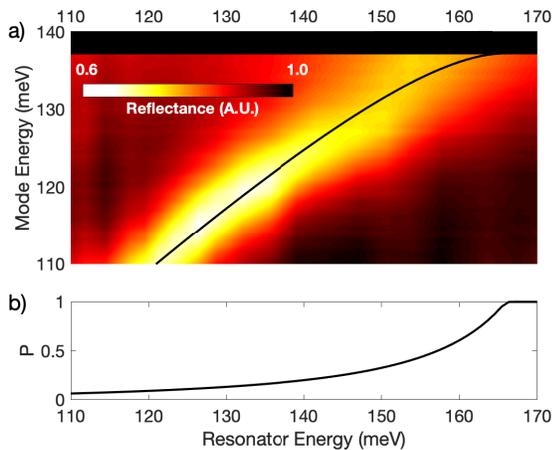}
\caption{\label{Figure_theory} 
{ \bf Calculation of the matter weight.} 
In panel (a) we plot the eigenmodes obtained by our theoretical model with parameters choose to fit the experimental reflectance data shown in the colorplot.
Above the ionization threshold the figure is uniformly black as the discrete modes used in the numerical diagonalization are too closely-spaced to be resolved.
In panel (b) we use the parameters extracted from panel a) to calculate the matter fraction $P$ of the discrete polariton mode.}
\end{center}
\end{figure}

Using the theory of Ref.~\cite{Cortese2019} such a hybrid discrete state can be written as a polariton whose electron density relative to the ground state one is given by
\begin{align}
\Delta{N}(z)&=P\left[  \lvert\psi^e(z)\rvert^2 -\lvert\psi^g(z)\rvert^2 \right],
\end{align}
where $P$, varying between $0$ and $1$, is the weight of the polaritonic matter component, $\psi^g(z)$ is the normalised wavefunction of an electron in its ground state,
and $\psi^e(z)$ is the wavefunction of a localised electronic state generated by the light-matter interaction.
In Fig.~\ref{Figure_theory}(a) we use the model of Ref.~\cite{Cortese2019} to reproduce the observed reflectivity spectrum and compare it to the experimental data. Using the parameters of such a simulation we calculated the matter weight $P$, which we plot in Fig.~\ref{Figure_theory}(b).
From such a figure we can see that the discrete resonance below the ionization threshold is clearly defined for non-vanishing values of $P$, demonstrating a substantial occupation of the light-generated electronic wavefunction $\psi^e(z)$.

With this experiment we demonstrated the possibility to strongly couple a ionizing transition to a photonic resonator.
As theoretically predicted, this results in a non-perturbative modification of the system's electronic structure. This leads to the appearance of a hybrid polaritonic excitation whose matter part is a bound state generated by the light-matter interaction. 
The experiment reported in this work concerns only the {\it optical} properties of the system. However, an immediate question concern the electronic transport: can these states conduct current?
If these bound states can indeed serve as a discrete and tunable current channels into bright states, it has been predicted that they could allow for dramatic increase in the efficiency of intersubband mid-infrared emitters ~\cite{DeLiberato2008b}. 
Initial work in this direction, albeit for a bound-to-bound transition, has been recently reported in mid-infrared detectors operating in the strong light-matter coupling regime~\cite{Vigneron2019}.

Beyond the boundary of mid-infrared optoelectronics, the possibility to tune material characteristics by coupling them to a microcavity photonic field has been much discussed lately, with particular attention for chemical and structural properties ~\cite{Galego2015,Ebbesen2016,Ruggenthaler2018}. Our results demonstrate this concept for semiconductor heterostructures, demonstrating that cavity quantum electrodynamics can be used as tool in quantum material engineering, not only shifting and hybridising resonances, but non-perturbatively and controllably changing their nature.

Finally, one tantalising feature of the discrete resonances we observe is their conceptual proximity with Cooper pairs, bound states of two electrons bound by the exchange of virtual phonons ~\cite{Cooper1956}. 
Theoretical proposals have appeared, showing that the critical temperature of a superconductor could be theoretically modified by coupling electrons through the exchange of virtual polaritons ~\cite{Laussy2010} or virtual cavity photons ~\cite{Ebbesen2016,Schlawin2019,Curtis2019}. 
And very recently, a first experimental result was reported~\cite{Thomas2019}.
Although the excitation we observed here is both electrically neutral and has a finite lifetime due to its polaritonic nature, it remains a first demonstration of the possibility to bound two particles not via Coulomb interactions but through the exchange of photonic excitations mediated by an judiciously engineered resonator.

\section{Acknowledgements}
S.D.L. is a Royal Society Research Fellow; R.C., J.M.M., G.B. and I.C. acknowledge partial financial support from the European Union FET-Open Grant MIR-BOSE 737017; R.C. and A.B. acknowledge partial financial support from the French National Research Agency (project ``IRENA''). This work was partly supported by the French RENATECH network.

\section*{Methods}

\subsection*{Sample growth and fabrication}
{
The samples were grown by molecular beam epitaxy (MBE) on semi-insulating 2-inch GaAs wafers. The growth starts with a 500-nm-thick Al$_{0.50}$Ga$_{0.50}$As stop layer, followed by 13 repetitions of GaAs QWs in Al$_{0.33}$Ga$_{0.67}$As barriers, and it ends with a 30-nm-thick GaAs cap layer. 
The two samples differ in QW width and introduced electronic doping, while the barriers are always 10-nm-thick. 
Sample HM4229 contains 4-nm-thick GaAs QWs ($L_{\mathrm{QW}}=4$ nm), while sample HM4230 contains 3.5-nm-thick QWs ($L_{\mathrm{QW}}=3.5$ nm).
Si doping is introduced in the center of the AlGaAs barriers as  $\delta$-doping, yielding a 2D electron density in the wells (measured at $300$K) of
$5\times 10^{12}$cm$^{-2}$ (HM4229) and $4.77\times 10^{12}$cm$^{-2}$ (HM4230), respectively.

The metal-metal resonators were realized by bonding the samples with Gold-gold thermocompression onto undoped GaAs wafers, following the procedure in  Ref.~\cite{Manceau2018}. After substrate removal, the resonators were defined with e-beam lithography followed by metal evaporation (Ti/Au) followed by lift-off.

\subsection*{Spectroscopic characterization}
The transmission spectra of the samples before wafer-bonding are acquired with standard multi-pass waveguide transmission measurements, after the samples have been metallized with Ti/Au on the top surface, and 45 degrees facets are shaped with mechanical polishing. The measurements are performed with an FTIR spectrometer, equipped with a thermal globar source, and a deuterated triglycine sulfate (DTGS) detector. 

The reflectivity measurements have been performed with a microscope connected to the FTIR. The microscope optics/mirrors operate in the mid-IR, the source is a thermal globar, while the detector is a liquid-nitrogen cooled Mercury-Cadmium-Telluride (MCT). Light is focused on the sample with a Cassegrain objective that operates  at a fixed incidence angle of incidence of $20^\circ \pm 5^\circ$. A very thin cryostat (Linkam LNP96) fits below the microscope objectives and permits to perform the reflectivity measurements at a temperature of 78K.}

\subsection*{Fitting procedure}
Both the resonances and the lifetimes showed in Fig.~\ref{Figure_exciton} have been obtained from the experimental reflectance data set for each metal stripe size, through an automatic multiple lorentzian peak fitting procedure. Same fitting model has been used for both doped and undoped samples' data.

In order to calculate the matter component $P$ use used the same procedure developed in Ref.~~\cite{Cortese2019}, using the doping and the ionization energy as adjustable parameters.

\bibliography{References}

\end{document}